# Laser Cooling of Germanium Semiconductor Nanocrystals

Manuchehr Ebrahimi[1], Wei Sun[2], Amr S. Helmy[1], Nazir P. Kherani[1,3]
[1]Department of Electrical and Computer Engineering, University of Toronto, Toronto, ON M5S 3G4 Canada
[2]Department of Chemistry, University of Toronto, Toronto, ON M5S 3H6 Canada
[3]Department of Materials Science and Engineering, University of Toronto, Toronto, ON M5S 3E4 Canada

Laser cooling of matter through anti-Stokes photoluminescence, where the emitted frequency of light exceeds that of the impinging laser by virtue of absorption of thermal vibrational energy, has been successfully realized in condensed media, and in particular with rare earth doped systems achieving sub-100K solid state optical refrigeration. Studies suggest that laser cooling in semiconductors has the potential of achieving temperatures down to ~10K and that its direct integration can usher unique high-performance nanostructured semiconductor devices. While laser cooling of nanostructured II-VI semiconductors has been reported recently, laser cooling of indirect bandgap semiconductors such as group IV silicon and germanium remains a major challenge. Here we report on the anomalous observation of dominant anti-Stokes photoluminescence in germanium nanocrystals. We attribute this result to the confluence of ultra-high purity nanocrystal germanium, generation of high density of electron-hole plasma, the inherent degeneracy of longitudinal and transverse optical phonons in non-polar indirect bandgap semiconductors, and commensurate spatial confinement effects. At high laser intensities, laser cooling with lattice temperature as low as ~50K is inferred.

Peter Pringsheim in 1929 proposed the idea of optical cooling of matter, reasoning the viability of continuous cooling of a fluorescent gas through a process called anti-Stokes photoluminescence where the luminescent frequency exceeds the frequency of the impinging optical field by virtue of absorption of thermal vibrational energy from the system [1]. Landau some two decades later confirmed the thermodynamic feasibility of anti-Stokes photoluminescent cooling, showing that the loss of entropy of the cooled medium is compensated by an increase in the entropy of the emitted light owing to loss of monochromaticity and omnidirectional emission *vis-à-vis* the incident light [2]. In the seventies Hänsch and Schawlow, and Wineland and Dehmelt, concurrently proposed laser cooling of atoms through the Doppler effect where absorption of red-detuned laser light followed by spontaneous emission at a higher frequency leads to a reduction in the translational velocity of the atoms [3,4]. Laser cooling of the gas phase has since advanced enormously, reaching micro-kelvin temperatures and enabling a manifold of applications including ultra-sensitive spectroscopy and study of Bose-Einstein condensates [5].

Laser cooling of solids in contrast has been a longer journey considering the challenges of mitigating non-radiative decay processes and increasing quantum efficiency of anti-Stokes photoluminescence. Kastler and Yatsiv in the fifties proposed the use of rare-earth doped transparent solids wherein the shielding of the optically active electrons by the outer shell electrons would limit the interaction with the host lattice and thus suppress non-radiative transitions [6,7]. Kushida and Geusic in 1968 reported the first evidence of cooling in an $Nd^{3+}$: YAG crystal [8] and then in 1995 Epstein and co-workers demonstrated net cooling in a $Yb^{3+}$-doped glass – made possible by the high material purity of the vitreous fluoride host, leveraging advances in the development of low-loss optical fibers in the near-infrared, and the high luminescence quantum efficiency of the radiative transitions [9]. With continual advances in material purity of rare earth doped systems, solid-state optical refrigeration is now in the sub-100 kelvin range.

Laser cooling of semiconductors in comparison has been an even more arduous trek [10-18]. Semiconductors are particularly attractive given the potential of achieving even lower temperatures of ~10 K and thus the opportunity to enhance the performance of various optoelectronic devices through direct integration. In contrast to rare-earth doped systems, where Boltzmann statistics leads to a significant reduction in population of the ground state manifold as temperatures drop below 100 K, semiconductors follow Fermi-Dirac statistics where the valence band remains populated all the way down to absolute zero temperatures and hence cooling transitions remain accessible. Early work on cooling of direct bandgap III-V GaAs was hampered by surface recombination losses and low external quantum efficiency due to its large optical index; in time these were substantially overcome through surface passivation and optical impedance matching techniques albeit large parasitic band tail absorption has prevented the attainment of net cooling. Laser cooling effects in spatially confined II-VI direct bandgap semiconductors have been reported recently where some of the key contributing factors are nano-length scale effects and low non-radiative surface recombination and Auger losses. This report has since sparked significant interest in optical cooling of a variety of direct bandgap semiconductor nanocrystal and nanostructure systems. However, laser cooling of indirect bandgap semiconductors such as germanium and silicon remains elusive.

Under normal conditions in most substances, optical excitation causes the local temperature to rise due to relaxation of high energy photogenerated carriers through lattice and electron-electron scattering leading to optical phonon emission. Raman spectroscopy, a common method used to study photon-phonon interactions, characterizes the photon scattering probability by the relation $W(\omega_i, \omega) = \frac{2\pi^2}{\hbar^2}\omega_i \omega N_i \frac{1}{4\pi}D(\omega)|M|^2$, where the scattered photon frequency $\omega = \omega_i \pm \omega_k$, the + and - symbols represent the anti-Stokes and Stokes transitions, respectively, $\omega_i$ the incident photon frequency, $\omega_k$ the phonon frequency, $N_i$ the number of incident photons, $D(\omega)$ the photonic density of states at the scattered photon frequency, and $M$ the matrix element of the scattering interaction.

The concept of laser cooling is illustrated in **Fig.1a**, where $E_{EX}$ and $K_BT$ are the excitation and the phonon energies, respectively, $E_{SPP}$ is the plasmon energy, and $E_{PL}$ is the photoluminescent energy. Laser [10-18] and Raman [19,20] cooling of doped and undoped direct bandgap semiconductor have been practically confirmed. Theoretically it has been shown that laser cooling in all semiconductors—direct and indirect bandgap materials—ought to be possible through engineering of the photonic density of states [21]. That is, the net radiative transition rate can be tailored considering that interband transition rate in semiconductors follows Fermi's golden rule, $W_{i \to f} = \frac{2\pi}{\hbar}|M(E_{PL})|^2 g(E_{PL})$, where $|M(E_{PL})|$ and $g(E_{PL})$ are the matrix element for radiative transition and joint density of state function, respectively, and noting that the latter term is a function of the density of electronic states and the density of photonic states. In this regard, it has been shown[8] that while in semiconductors at T > 150K and low carrier concentrations Auger process remains the limiting factor for laser cooling, at high carrier concentrations there exists an optimal cooling range of 50K to 120K where Auger recombination is low enough and radiative recombination is bimolecular but short of the saturation carrier concentration above which the radiative process becomes linear (i.e., dependent only on the hole concentration *N* rather than both electron and hole concentrations *N²*).

Herein, we find that net laser cooling is possible in ultra-high purity nanocrystal germanium via resonant photon-phonon coupling upon generation of high-density electron-hole plasma. We present experimental Raman spectra showing that the intensity of Stokes and anti-Stokes from germanium nanocrystals (ncGe) with diameters ranging from approximately 16 nm to 30 nm can be inverted with increasing laser power density using 785 nm (1.58 eV) wavelength light.

The ncGe nanoparticles were fabricated using an organic-free process entailing thermal disproportionation of GeO prepared from thermally induced dehydration of Ge(OH)$_2$ [22]. We posit that the high purity of ncGe and associated amorphous phase and any residual oxygen and hydrogen play an important passivation role that contributes to this first evidence of laser cooling in indirect gap semiconductor nanocrystal.

Two types of experiments were carried out to investigate laser cooling of ncGe nanoparticles; first, where ncGe nanoparticles are dispersed in isopropyl alcohol (IPA) solution in a Teflon capillary tube, and second, where few droplets of ncGe in hexanol solution is dispensed on a quartz microscope slide resulting in dried ncGe powder. Considering that the Stokes and anti-Stokes measurements were not simultaneous, a series of Stokes and anti-Stokes measurements were carried out in alternating sequence on fresh nGe in IPA samples thus ensuring the absence of any systematic effects. Further, a series of background measurements were also carried out on IPA samples only (*san* ncGe). Similarly, a series of Stokes and anti-Stokes measurements were carried out on ncGe powder such that each measurement was carried out on a fresh region of the sample. We note that all ncGe samples were stored in IPA and hexanol to mitigate any oxidation effects. Experimental details are illustrated in **Fig. 1 (b)**, **(c)** and **(d)**, and further information is provided in the Supplementary.

Typical Anti-Stokes and Stokes photoluminescence spectra, due to 785 nm laser excitation at room temperature, for nanocrystal germanium, both 25nm ncGe dispersed in IPA solution and 25nm ncGe powder in hexanol dispersed on a quartz substrate, are shown in **Fig. 2(a)** and **2(b)**, respectively. Considering that the hexanol evaporates rapidly the resulting sample is denoted powder ncGe. Both ncGe sample sets exhibit anti-Stokes intensity exceeding the Stokes intensity at $\omega_k \approx$ 292 cm$^{-1}$ which is the first order Raman active optical phonon mode of nanocrystalline germanium with ~37 meV energy. It is also observed that while the Raman Stokes and anti-Stokes signal intensities associated with ncGe in IPA (**Fig. (2a)**) are anomalous at ~292 cm$^{-1}$, the Raman peaks for IPA at 200 cm$^{-1}$ remain normal.

A series of Stokes and anti-Stokes spectra were obtained for ncGe in IPA and ncGe powder as a function of laser power ranging from 0.1% (0.015 mW) to 100% (15 mW) at 785 nm wavelength. These spectra were then analyzed for the anti-Stokes ($I_A$) to Stokes ($I_S$) signal strength ratio as a function of the laser power, as shown in **Fig. 3(a)**, where the dashed trend lines are a guide to the eye. Under low laser intensity ≲ 10% (1.5 mW) illumination, the anti-Stokes to Stokes signal strength ratios are typical of less than unity, while at higher laser intensities we observe anomalous anti-Stokes to Stokes ratios exceeding unity. We further note that the general trend for ncGe in IPA and ncGe powder are similar, suggesting that the simpler dispensation of ncGe powder on quartz is a viable means of studying laser cooling effects in ncGe samples. Three sets of ncGe powder samples with the following diameters, 16nm ± 4nm, 25nm ± 5nm, and 30 nm ± 7nm size (where ± s indicates estimated standard deviation of the distribution in nanoparticle size) were studied. We note that while there is scatter in the data, which is not surprising considering that the Stokes and anti-Stokes measurements were each carried out sequentially over a fresh region of the sample, the similarity in the trends of $I_A/I_S$ with laser power is striking.

Assuming thermal equilibrium we inferred the lattice temperature of the nanoparticles on the basis of the ratio of Stokes and anti-Stokes signal strengths for the characteristic Raman band for Ge using the Boltzmann-Einstein distribution:

$$\frac{I_A(\omega_k)}{I_S(\omega_k)} = \left(\frac{\omega_i + \omega_k}{\omega_i - \omega_k}\right)^4 e^{\left(-\frac{\hbar\omega_k}{KT}\right)}$$

where $I_A$ and $I_S$ are the intensities of anti-Stokes and Stokes bands with vibrational frequencies of $+\omega_k$ and $-\omega_k$, respectively, $\omega_i$ is the excitation frequency, $T$ is the lattice temperature, and $K$ and $\hbar$ are the Boltzmann and Planck constants, respectively [23,24]. The inferred lattice temperatures are presented as a function of the laser power in **Fig. 4**. It is observed that the lattice temperature increases with increasing power up to a laser intensity of ~10%; in fact, the inferred lattice temperatures are remarkably high at 10% power level and in particular increase with decreasing ncGe nanoparticle diameter. At laser intensities above 10% laser power the lattice temperature drops below ~100K for the most part, while minimum lattice temperatures of ~50$K$ are observed at 25% (~4 mW) power. It is interesting to observe that at laser powers exceeding 25% (~4mW) the inferred lattice temperatures increase with increasing power, which may indicate that the system is entering the saturation carrier density regime where the radiative transition rate assumes a linear dependence on $N$.

We now examine the physical processes at play in the nanocrystal germanium system that under elevated carrier and phonon densities give rise to the observed laser. In most materials, the total thermal conductivity, $\overleftrightarrow{k} = \overleftrightarrow{k}_e + \overleftrightarrow{k}_L$, is the superposition of electronic ($\overleftrightarrow{k}_e$) and lattice ($\overleftrightarrow{k}_L$) contributions, where for the semiconductor, $\overleftrightarrow{k}_e = \left(\frac{35}{2}\right)\left(\frac{K_B}{e}\right)^2 \sigma T$ and $\overleftrightarrow{k}_L = \frac{C_p \bar{v}_q \Lambda_{ph}}{3}$ with the parameters $\sigma, C_p, \bar{v}_q, \Lambda_{ph}$ defined as electrical conductivity, heat capacity, average phonon velocity and phonon mean free path, respectively. In semiconductors at low carrier densities heat is dominantly carried by the lattice, i.e., principally the acoustic phonons, while at high carrier densities both lattice and electronic contributions play a role. It is interesting to note that the slope in **Fig. 3** is indicative of the thermal conductivity of the sample and as such the principal slopes at low and high laser power intensities are reflective of dominant processes at play which result in sample heating and cooling, respectively. Specifically, the thermal conductivity of the sample is $k = \frac{Lq}{\Delta T}$ where $L$ represents the sample thickness and $q = Q/A$ is the heat flux, with $Q$ being the conducted heat and $A$ the cross-sectional area corresponding to the laser spot size, and $\Delta T$ is the change in the local temperature due to the heat flux in the specimen. Recognizing that the local temperature is proportional to the ratio of the anti-Stokes to Stokes intensities, and that $L/A$ is essentially a constant, thus the slope in **Fig. 3** is inversely proportional to the thermal conductivity of the specimen [25].

At low laser intensity we observe high lattice temperatures implying significant heat accumulation which is consistent with a relatively low thermal conductivity of the sample. This is further understood by noting that at temperatures above the Debye temperature ($T \gg \Theta$) heat capacity is constant ($C_p = 3NK_B$ where $N$ is Avogadro's number) and independent of temperature while the phonon mean free path $\Lambda_{ph}$ diminishes and hence leads to a low thermal conductivity $\overleftrightarrow{k} \approx \overleftrightarrow{k}_L$. The mean free path depends on phonon-phonon interaction and accordingly is a function of the phonon density $n(\mathbf{q})$ which is given by the Bose-Einstein expression. The phonon density, $n(\mathbf{q}) = \left(e^{\hbar\omega(\mathbf{q})/K_B T} - 1\right)^{-1}$, at high temperature reduces to $n(\mathbf{q}) \approx K_B T/\hbar\omega(\mathbf{q})$. Higher phonon density at higher temperature promotes phonon-phonon scattering processes which lowers the mean free path and hence the thermal conductivity. Also, higher temperatures give rise to higher $\mathbf{q}$ values which lead to Umklapp processes that limit thermal conductivity in crystalline materials. Additionally, phonon-boundary scattering has a bearing on the phonon mean free path as the nanocrystal size becomes comparable to $\Lambda_{ph}$.

At high laser intensity we observe low lattice temperatures and correspondingly the slope in **Fig. 3** indicates high thermal conductivity of the sample. This is appreciated by noting that while at low temperatures lattice thermal conductivity has a $T^3$ dependence, given $C_p \propto T^3$, the electronic thermal conductivity under high laser intensity becomes the dominant contributor given its dependence on carrier density. This in turn promotes electron-phonon scattering whereas the system transitions (knee near 5-10% power in **Fig. 3**) to elevated carrier and phonon densities, the probability of phonon absorption begins to dominate over the phonon emission process. This is further appreciated upon examining **Fig. 4** where we observe an increase in the inferred lattice temperature up to the knee point of ~10% power prior to the cooling effect beginning to dominate. Expectedly the maximum inferred lattice temperature is a function of the nanocrystal size, where the smaller nanocrystals exhibit higher peak temperatures considering constrained heat dissipation owing to the smaller surface area.

Examining this further, we note that conservation of energy and momentum require that $E_f = E_i \pm \hbar\omega_q$ and $\boldsymbol{k}_f = \boldsymbol{k}_i \pm \boldsymbol{q}$, where +/- denotes absorption/emission of a phonon of energy $\hbar\omega_q$ in the scattering of an electron by a phonon of wavevector $q$; $E_i$, $E_f$, $k_i$ and $k_f$ represent the initial and final electron energy and initial and final electron wavevector, respectively, and $\omega_q$ is the phonon frequency. Now the electron transition probability from an initial state $i$ to a final state $f$ is a function of the availability of the final electronic states and the probability of phonon emission or absorption. Considering that all the electronic states in the conduction band are essentially accessible, the availability of the final states is simply the density of the final electron states times the probability that the final state is unoccupied (viz., unity); that is, $\rho(E_f^\pm) = \frac{(2m^*)^{\frac{3}{2}}}{2\pi^2\hbar^3}\sqrt{E_i \pm \hbar\omega_q}$. The probability of phonon absorption and emission are proportional to the electron-phonon coupling strength $G(q)$ and the phonon density for absorption $n(q)$ and the phonon density for emission $[1+n(q)]$, where $n(q) = \left(e^{\hbar\omega_q/KT} - 1\right)^{-1}$. Combining the absorption and emission terms and summing over all the final states yields the scattering probability ($1/\tau_c$),

$$\frac{1}{\tau_c} \sim \sum_q G(q)\left[\rho(E_f^+)n(q) + \rho(E_f^-)(1 + n(q))\right]$$

where the first and second terms in the square bracket correspond to phonon absorption and emission, respectively. The electron-phonon coupling coefficient $G(q)$ in non-polar crystals is dominated by the deformation-potential coupling mechanism, and accordingly both phonon absorption and emission factors ($G(q)n(q)$ and $G(q)(1+n(q))$) are independent of $q$ for the LA branch. For the acoustic phonon, the primary scattering comes from the long wavelength acoustic phonons [26,27], so $\omega \sim q$ and $G(q) \sim q$. This further simplifies the scattering probability where it is only a function of temperature and the density of final electron states for either phonon absorption or emission, $\frac{1}{\tau_c} \sim K_B T \sum [\rho(E_f^+) + \rho(E_f^-)]$, that is, independent of the phonon. Further, as determined by the Lyddane-Sachs-Teller relationship for negligibly damped purely covalent crystals of the group IV elements, the LO and TO phonon modes are degenerate in Ge – having the same frequency at the Brillouin zone center. Thus, the LO modes can couple to the plasmon waves [28], thus enhancing electron-phonon interaction leading to phonon absorption, plasmon emission and light scattering by plasmon [29], and hence laser cooling [30].

Nanostructures such as Ge nanowires have been reported to exhibit optical phonon peak shifts which are dependent on the laser excitation power but independent of wavelength [31-33]. Silicon and germanium, non-polar crystals with indirect bandgaps, have been generally deemed to be inaccessible to laser cooling [21]. However, in nanostructures the distinction between direct and indirect bandgap fades as the electron and hole wave functions spread in momentum space, breaking the usual crystal momentum selection rules [34]. Overlapping of electron and hole wave functions accompanying reduction in size increases the coupling between their transition matrix elements and thus allows zero-phonon transitions where the amplitude of this transition is strongly size dependent. As in other reports[35] we also observe (**Fig. 5**) downshifting and broadening of the Stokes and anti-Stokes peaks for nanostructures smaller than 300Å. Additionally, we note that ordinarily unpassivated surface atoms in ncGe would tend to weaken the oscillation strength and hence surface modes would appear at low frequencies; in fact, these have been reported [36] to be below $50 \, cm^{-1}$ with one exception at $260 \, cm^{-1}$. Moreover, Lamb modes, also known as breathing modes for spherical systems which tend to be inversely proportional to the nanoparticle diameter $(1/d)$, have Raman peaks at low frequencies of $< 100 \, cm^{-1}$. In contrast, in our high purity and well passivated ncGe the Raman peaks are observed at $\sim 292 \, cm^{-1}$ (~36 meV) – the principal transverse optical mode of ncGe.

We further note that extensive Coulombic interactions within the spatially confined structures contribute to various charge scattering processes. Carrier multiplication (CM) is a charge scattering process that reduces the amount of energy lost to local heating by generating additional free carriers. These free carriers can then participate in band-to-band transitions and thus energy is radiatively removed from the system through the emission of lower energy photons. Indeed, the CM process has been observed in ncGe [37] where interestingly charge multiplication has been detected below the onset of twice the Ge bandgap. In the present experiment, Raman cooling was observed with a laser wavelength of 785 nm or ~1.58 eV, photon energy well above the Ge bandgap. In this regard, it is worth mentioning that although photoluminescence relies on excited electronic states corresponding to eigenstates of a physical system, Raman scattering can transpire via pseudo or virtual states of a system. The presence of pseudo resonant states in plasmonic crystals has been reported [38] and the experimental results presented herein are potentially suggestive of this phenomenon.

We lastly mention other potential phenomena present within the present ncGe system and the need to closely examine their interplay and role in the observed Raman cooling. The generation of hot carriers, for example through charge multiplication, may lead to ballistic conditions which may give rise to perturbations in the dielectric function of the free electron gas in a bounded system and thus lead to longitudinal plasmon waves otherwise known as Langmuir waves. Considering a spatially confined system with length scales less than the Debye length, this may lead to a dispersionless electron dispersion relationship with implications thereof. Further, under high intensity laser illumination and accompanying electron plasma, the role of ion acoustic waves needs to be understood. Moreover, further compositional, structural, optical/photonic and electrical characterizations are required to completely understand the nanochemistry of these group IV nanocrystal systems.

In summary, we report first experimental evidence of Raman laser cooling in an indirect bandgap semiconductor. The high purity nanocrystal germanium system exhibits optical cooling under high intensity laser illumination. Raman cooling measurements are reported for both liquid and dry samples. We identify a range of key potential phenomena that contribute to the observed anomalous dominant anti-Stokes emission.

**Conflicts of interest**

There are no conflicts of interest.

**Acknowledgements**

The authors acknowledge the support of the Natural Sciences and Engineering Research Council of Canada (NSERC).

**Author Contributions**

Systematic experiments were designed and carried out by Manuchehr Ebrahimi, followed by detailed analysis of the results and preparation of the first draft of the manuscript both of which were also done by him. Manuchehr Ebrahimi discovered the anomaly and recognizing its potential significance developed the physical basis for the anomalous observation. The final manuscript was jointly revised by Manuchehr Ebrahimi and Nazir P. Kherani. The samples were prepared by Wei Sun in Geoffrey A. Ozin's Nanomaterials Chemistry group. The Raman measurements were carried out by Manuchehr Ebrahimi using the Horiba Raman Facility in Amr Helmy's group.

**Data Availability**

All data will be available upon reasonable request.

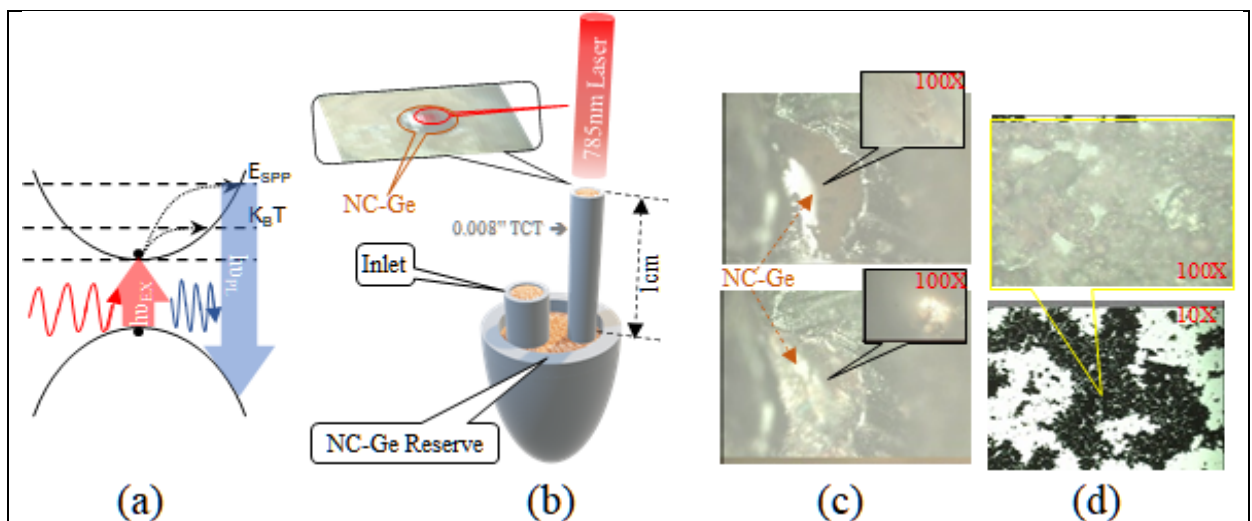

**Fig. 1**. (**a**) An illustration of Raman laser cooling processes under high laser power density; $h\nu_{EX}$ and $h\nu_{PL}$ are the excitation and photoluminescent energies, $K_BT$ and $E_{SPP}$ are the phonon and plasmon energies. (**b**) An illustration of the Raman laser cooling measurement method for nc-Ge in isopropyl alcohol (IPA) wherein laser illumination is focused on the nc-Ge in IPA within the Teflon capillary tube (TCT); the nc-Ge in IPA within the TCT is periodically refreshed. (**c**) Photographs of the facet/plane at the top of the TCT filled with nc-Ge in IPA under 100x magnification before (top photograph) and after (bottom photograph) exposure to 785nm laser, illustrating 'film' formation on the top. (**d**) Images of nc-Ge powder on a microscope slide at 10x and 100x magnification prior to exposure to 785nm laser excitation.

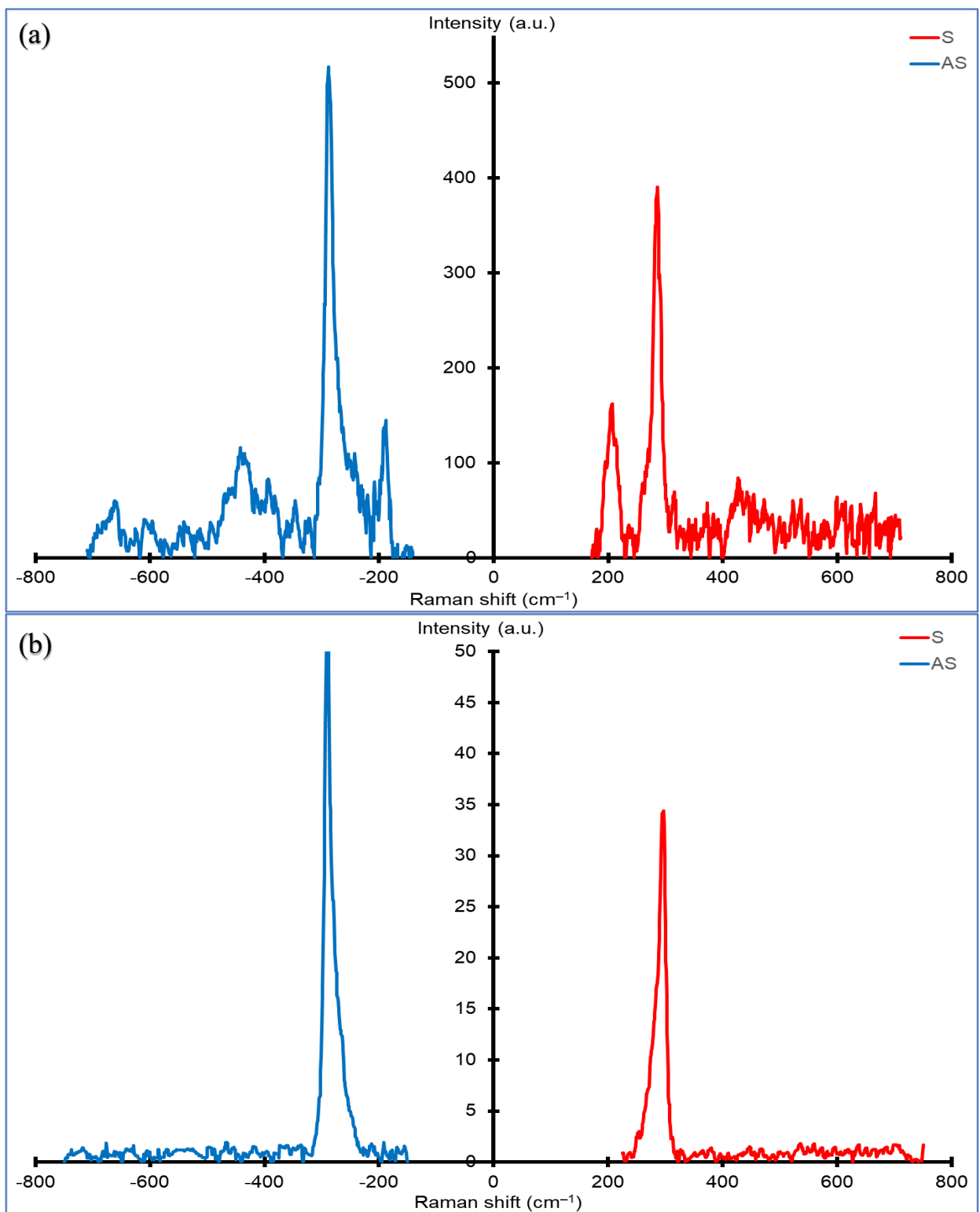

**Fig. 2**. Raman spectra of nanocrystals of germanium (ncGe) illuminated by 785 nm laser light. (**a**) Stokes and anti-Stokes peaks from 25nm ncGe dispersed in IPA solution when exposed to 15 mW (100%) laser power. (**b**) Stokes and anti-Stokes peaks from 25nm ncGe powder exposed to 3.8 mW (25%) laser power.

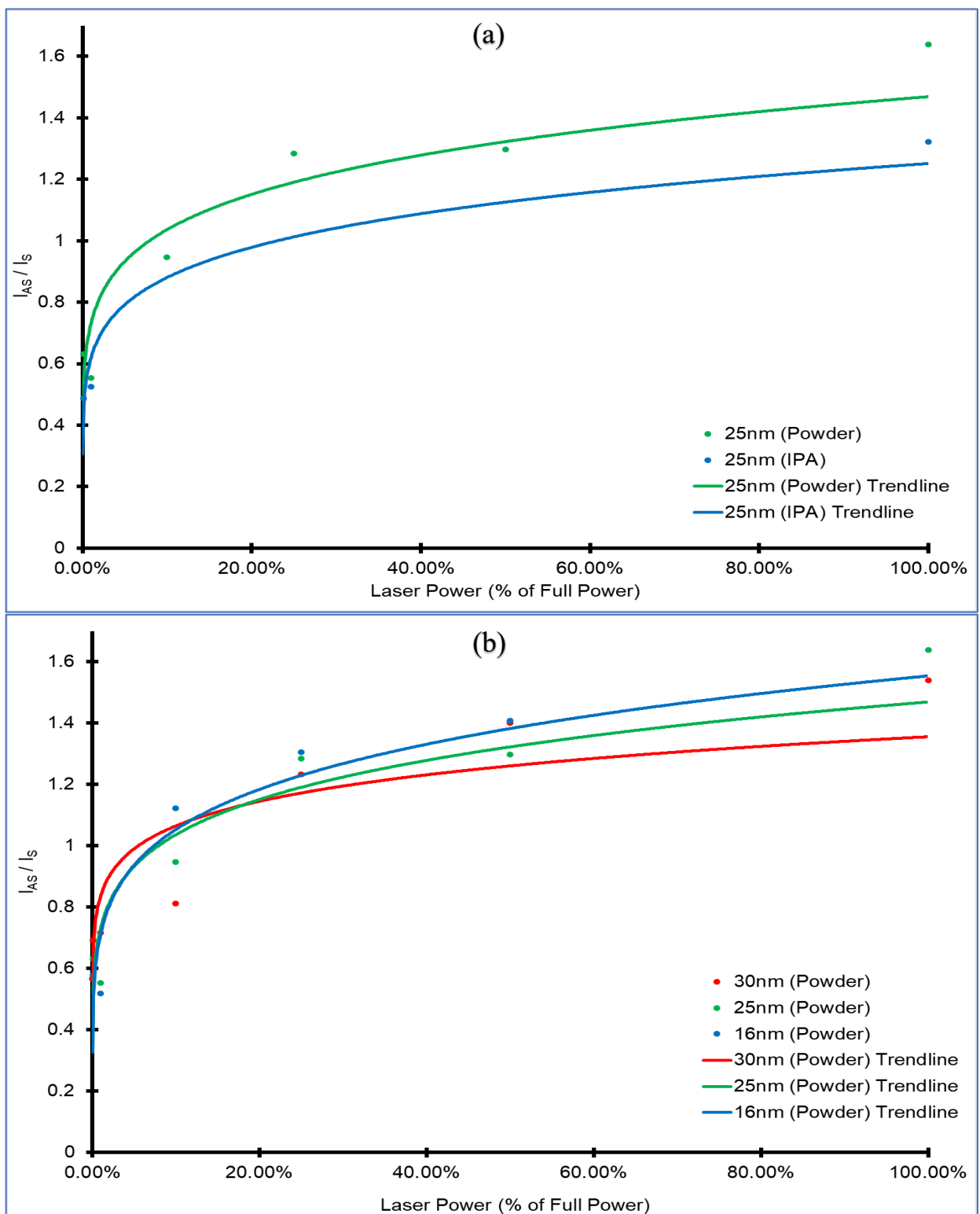

**Fig. 3**. Ratio of anti-Stokes to Stokes peak intensities, representative of the thermal conductivity conductivity of the ncGe as a function of laser power. (**a**) Variation in thermal conductivity with laser power for 25nm ncGe in IPA solution and 25nm ncGe powder. (**b**) Variation in thermal conductivity with laser power for 16nm, 25nm and 30nm ncGe powder.

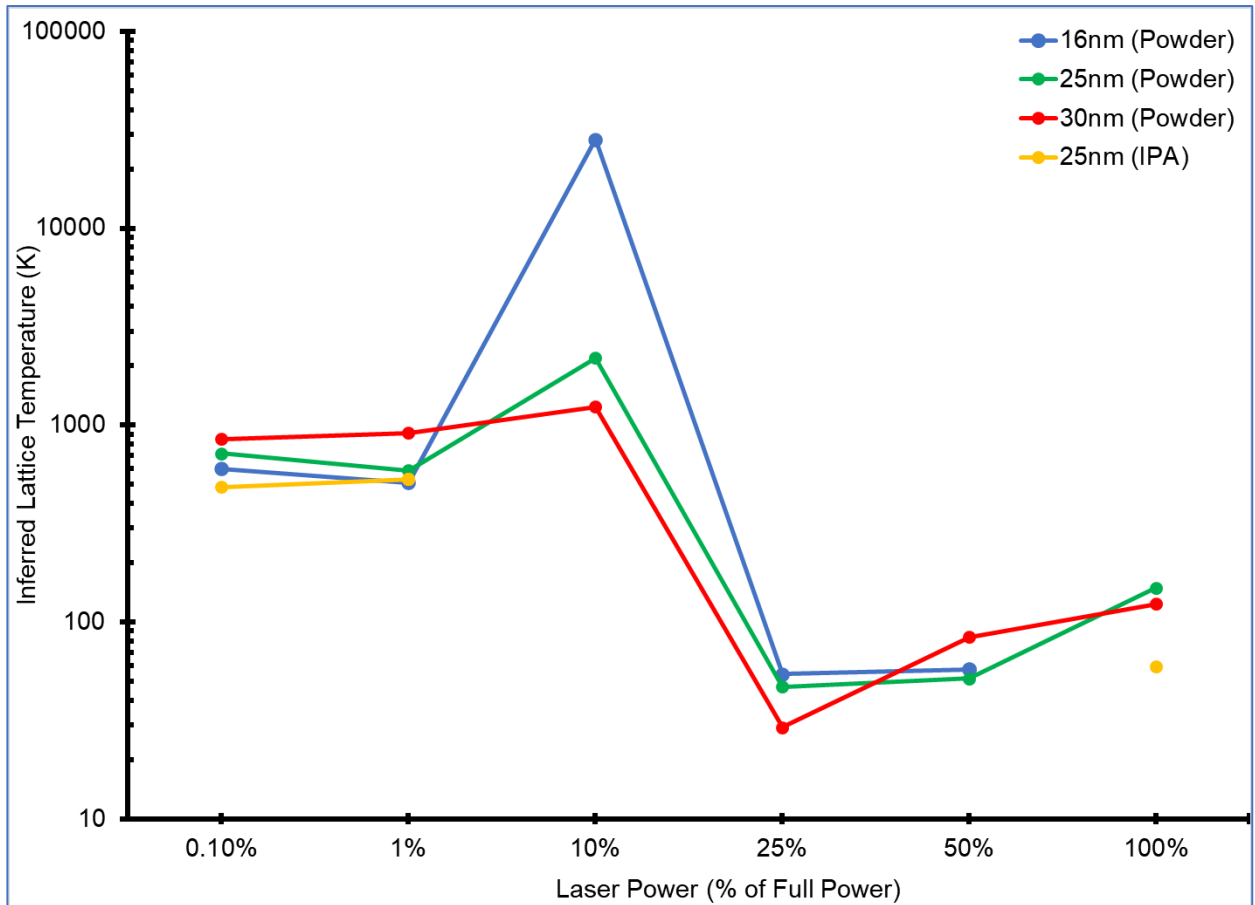

**Fig. 4**. Variation in inferred lattice temperature with laser power for 16nm, 25nm and 30nm ncGe powder. The lines are a guide to the eye for the general trend. Inferred lattice temperatures for 25nm ncGe in IPA are also included.

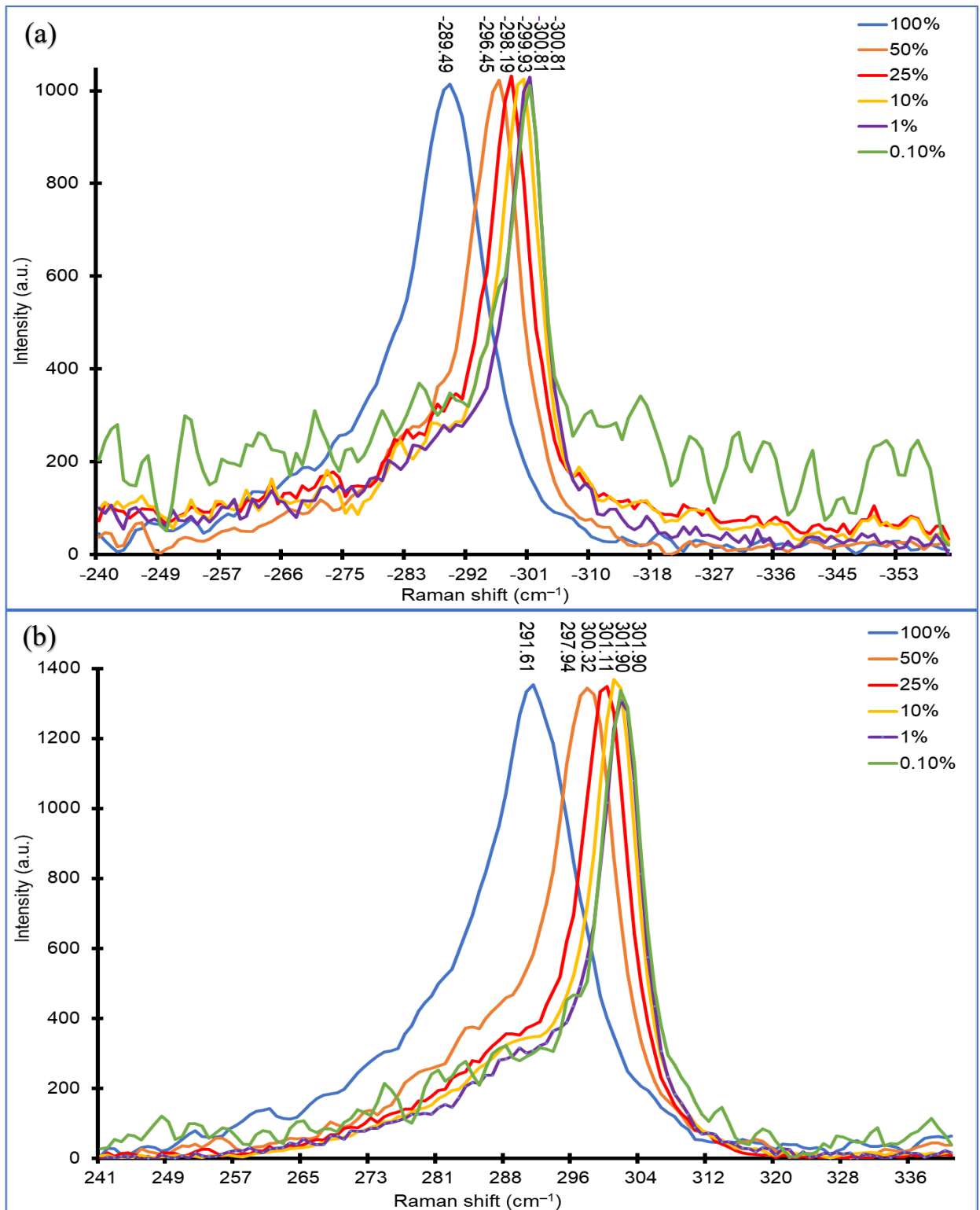

**Fig. 5**. Raman peak shifts for 25nm ncGe powder with laser power. (**a**) Normalized anti-Stokes and (**b**) normalized Stokes spectra for power ranging from 0.015 mW (0.1%) to 15 mW (100%) laser power (as indicated in the legend).